\documentclass[a4paper,12pt]{article}
\pdfoutput=1
\usepackage{graphicx,rotating,hyperref,slashed,amsmath,xcolor,amssymb,amsfonts,colortbl,expdlist}
\makeatletter
\def\citep{\cite}
\usepackage{soul}
\usepackage{blkarray}
\usepackage{multirow,multicol}
\usepackage{dsfont}
\usepackage{footnote}

\hypersetup{colorlinks,bookmarksopen,bookmarksnumbered,
linkcolor=blus,pdfstartview=FitH,urlcolor=rossos,citecolor=verde}
\allowdisplaybreaks

\makeatletter
\usepackage[natbibapa]{apacite}

\AtBeginDocument{%
  
}

\def\lsim{\mathrel{\rlap{\lower3pt\hbox{\hskip0pt$\sim$}}
   \raise1pt\hbox{$<$}}}         
\def\gsim{\mathrel{\rlap{\lower4pt\hbox{\hskip1pt$\sim$}}
   \raise1pt\hbox{$>$}}}         

\newcommand{\mio}[1]{}

\newcommand{\fig}[1]{~\ref{fig:#1}}

\definecolor{Gray}{gray}{0.95}

\definecolor{rosino}{cmyk}{0,0.05,0.05,0.02}
\definecolor{celestino}{cmyk}{0.05,0,0,0.01}

\usepackage{multicol}
\usepackage{color}
\definecolor{rosso}{cmyk}{0,1,1,0.4}
\definecolor{rossos}{cmyk}{0,1,1,0.55}
\definecolor{rossoc}{cmyk}{0,1,1,0.2}
\definecolor{blu}{cmyk}{1,1,0,0.3}
\definecolor{blus}{cmyk}{1,1,0,0.6}
\definecolor{bluc}{cmyk}{1,1,0,0.1}
\definecolor{verde}{cmyk}{0.92,0,0.59,0.25}
\definecolor{verdec}{cmyk}{0.92,0,0.59,0.15}
\definecolor{verdes}{cmyk}{0.92,0,0.59,0.4}

\newcommand{\eq}[1]{~{\rm (\ref{eq:#1})}}

\def\circa#1{\,\raise.3ex\hbox{$#1$\kern-.75em\lower1ex\hbox{$\sim$}}\,}

\newcommand{\beq}{\begin{equation}}
\newcommand{\eeq}{\end{equation}}

\newcommand{\bea}{\begin{eqnarray}}
\newcommand{\eea}{\end{eqnarray}}
\newcommand{\be}{\begin{equation}}
\newcommand{\ee}{\end{equation}}
\font\tenrsfs=rsfs10 at 12pt
\font\sevenrsfs=rsfs7 at 9pt
\font\fiversfs=rsfs5
\newfam\rsfsfam
\textfont\rsfsfam=\tenrsfs
\scriptfont\rsfsfam=\sevenrsfs
\scriptscriptfont\rsfsfam=\fiversfs
\def\mathscr#1{{\fam\rsfsfam\relax#1}}

\def\circa#1{\,\raise.3ex\hbox{$#1$\kern-.75em\lower1ex\hbox{$\sim$}}\,}
\makeatletter

\def\hhref#1{\href{http://arxiv.org/abs/#1}{arXiv:#1}} 

\newcommand{\doi}[1]{\href{http://dx.doi.org/#1}{[link]}}

\def\hhref#1{\href{http://arxiv.org/abs/#1}{arXiv:#1}} 
 
\def\art{\@ifnextchar[{\eart}{\oart}}
\def\eart[#1]#2#3#4#5#6{{\rm #2}, {\em #3 \bf #4} {\rm (#6) #5} ({\em #1})}

\def\article{\@ifnextchar[{\earticle}{\oarticle}}
\def\oarticle#1#2#3#4#5#6{{\rm #1}, {\em ``#6''}, {\rm #2 #3 (#5) #4}}
\def\earticle[#1]#2#3#4#5#6#7{{\rm #2}, {\em ``#7''}, {\rm #3 #4 (#6) #5}  [\hhref{#1}]}
\def\hepart[#1]#2{{\rm #2, \em#1}}
\def\heparticle[#1]#2#3{#2, {\em ``#3''} [\hhref{#1}]}
\def\inspire{{\sc InSpire} }

\newcounter{alphaequation}[equation]
\def\thealphaequation{\theequation\hbox to
0.6em{\hfil\alph{alphaequation}\hfil}}
\def\eqnsystem#1{
\def\@eqnnum{{\rm (\thealphaequation)}}
\def\@@eqncr{\let\@tempa\relax \ifcase\@eqcnt \def\@tempa{& & &} \or
  \def\@tempa{& &}\or \def\@tempa{&}\fi\@tempa
  \if@eqnsw\@eqnnum\refstepcounter{alphaequation}\fi
\global\@eqnswtrue\global\@eqcnt=0\cr}
\refstepcounter{equation} \let\@currentlabel\theequation \def\@tempb{#1}
\ifx\@tempb\empty\else\label{#1}\fi
\refstepcounter{alphaequation}
\let\@currentlabel\thealphaequation
\global\@eqnswtrue\global\@eqcnt=0 \tabskip\@centering\let\\=\@eqncr
$$\halign to \displaywidth\bgroup \@eqnsel\hskip\@centering
$\displaystyle\tabskip\z@{##}$&\global\@eqcnt\@ne
\hskip2\arraycolsep\hfil${##}$\hfil& \global\@eqcnt\tw@\hskip2\arraycolsep
$\displaystyle\tabskip\z@{##}$\hfil
\tabskip\@centering&\llap{##}\tabskip\z@\cr}
\def\endeqnsystem{\@@eqncr\egroup$$\global\@ignoretrue} \makeatother

\definecolor{fiorentina}{rgb}{.5,0,.5}

\definecolor{rossoCP3}{cmyk}{0,.88,.77,.40}

\begin{document}

\vspace{2cm}

\begin{center}
\boldmath

{\textbf{\LARGE\color{rossoCP3} Bibliometrics for collaboration works}}
\unboldmath

\bigskip\bigskip

\vspace{0.1truecm}

{\large\bf Paolo Rossi$^{a}$, Alessandro Strumia$^{a}$, and Riccardo Torre$^{b,c}$}
 \\[8mm]
{\it $^a$ Dipartimento di Fisica dell'Universit{\`a} di Pisa, Italy}\\[1mm]
{\it $^b$ CERN, Theory Division, Geneva, Switzerland}\\[1mm]
{\it $^c$ INFN, sezione di Genova, Italy}\\[1mm]

\vspace{1cm}

\thispagestyle{empty}
{\large\bf\color{blus} Abstract}
\begin{quote}
\large
An important issue in bibliometrics is the weighing of co-authorship in the production of scientific collaborations, which are becoming the standard modality of research activity in many disciplines.
The problem is especially relevant in the field of high-energy physics, where collaborations reach 3000  authors, but it can no longer be ignored also in other domains, like medicine or biology.
We present theoretical and numerical arguments in favour of weighing
the individual contributions as $1/N_{\rm aut}^\alpha$
where $N_{\rm aut}$ is the number of co-authors. 
When counting citations we suggest the exponent $\alpha\approx 1$, that corresponds to fractional counting.
When counting the number of papers we suggest $\alpha \approx 1/3 - 1/2$, with the former (latter) value more appropriate for larger (smaller) collaborations.
We expect and verify that the $h$ index scales  as the square root of the average number of co-authors,
and define a fractionalized $h$ index that does not scale with collaboration size.
\end{quote}
\thispagestyle{empty}
\end{center}

\setcounter{footnote}{0}

\setcounter{page}{1}
\setcounter{footnote}{0}

\newpage

\section{Introduction}
In many research fields, scientific collaboration has become the standard way of operating, and moreover, due to the increasing complexity of the problems to be faced, the number of  scientists with different competences involved in each single collaboration is  increasing. In the extreme case of high energy physics numbers have already reached the four-digit level, but in many other domains, like medicine or biology, it is not unusual to find two-digit collaborations.

In the context of bibliometrics this hyper-authorship phenomenon poses a very important question, concerning the individual degree of property that must be assigned to the authors of a common scientific article, both concerning the paper itself and the citations it receives.
It is rather clear that attributing the full credit of a paper to each of the authors  
 is  mystifying and 
{(if adopted by policymakers) }
 tends to encourage fictitious collaborations, because of the obvious competitive advantage resulting from the much larger number of articles that a collaboration may produce in the same amount of time in comparison with an isolated author. Moreover, also the number of citations received is strongly correlated with the typical dimensions of the collaborations operating in a given field of research.

Fractional counting of papers and citations could be a solution to this issue. Fractional counting has been extensively discussed in the literature. For instance, it has been considered in the context of metrics and rankings by \cite{DBLP:journals/jasis/Hooydonk97,Egghe08a,DBLP:journals/corr/abs-1007-4749,DBLP:journals/corr/abs-1006-2896,DBLP:journals/corr/abs-1106-0114,DBLP:journals/jasis/LeydesdorffS11,DBLP:journals/jasis/LeydesdorffB11,DBLP:journals/joi/AksnesSG12,DBLP:journals/aslib/Rousseau14,DBLP:journals/joi/BouyssouM16,Strumia:2018zgi}, and in the context of constructing research networks by \cite{DBLP:journals/joi/Perianes-Rodriguez16,DBLP:journals/joi/LeydesdorffP17}. 
Fractional counting gives an intensive quantity: this means, for example, that
the total index of the European Union is the sum of its members, unlike
what happens if full counting is adopted (see e.g.\ the discussions by \cite{DBLP:journals/scientometrics/GauffriauLMRI08,Strumia:2018zgi,DBLP:journals/joi/Gauffriau17}, and \cite{DBLP:journals/joi/WaltmanE15}).

However, the choice of fractionally counting papers by 
attributing a $1/N_{\rm aut}$ weight to each of the $N_{\rm aut}$ co-authors of a paper would imply a strong penalty for authors beloinging to large collaborations. As we will show in the following, see for instance fig.\fig{MeanNcitNaut}, this can be seen from the strong dependence of the fractionally counted number of papers on the collaboration size, which implies a strong reduction of the fractionally counted number of papers for collaborations with many co-authors. 
Indeed typically $N_{\rm aut}$ tend to produce less than $N_{\rm pap}=N_{\rm aut}$ papers 
in the same time in which a single author produces a single paper. 
While it is clear that from a pure research perspective what matters is the scientific impact, 
that cannot be quantified trough the publication frequency, there are still examples of policymakers that consider the number of papers as a simple relevant indicator and use it for the evaluation of scientists.\footnote{As an example, the Italian Ministry for Research poses lower thresholds in the publication frequency to access to professorship positions.}
While full counting of papers penalises authors working in small collaborations
(for instance small experiments in fundamental physics),
full fractional counting of papers, if adopted by policymakers, would on the contrary discourage large collaborations, 
which are a necessary endeavour in modern research.
This problem needs therefore a non-subjective solution, namely finding which 
compromise between these two extreme choices
gives bibliometric indicators that are as independent as possible from the size of collaborations and authors groups.

This general principle will give different answers when applied to different indicators:
counting of papers, of citations, $h$-index, etc.

The aim of the present paper is to find fractional counting algorithms that do not scale with collaboration size. This has a two-fold advantage: on the one hand it ensures that neither authors belonging to large collaborations, nor single authors nor authors working in small groups are favoured or disfavoured, when bibliometric indicators are is used in their evaluation, for their choice of carrying out their research in large versus small groups. 
On the other hand this allows to better quantify and qualify the bibliometric output of large collaborations in comparison with small groups of authors (or even single authors).


The bibliometric literature documents that
collaboration papers tend to have
higher impact than single-authored papers.
\cite{Beaver} studied the field of physics, finding that co-authored research tends to be of higher quality than
solo research.
\cite{Bordons1993} studied Spanish publications in
pharmacology and pharmacy
finding that internationally co-authored documents have higher impact 
than the remaining collaborative documents or non-collaborative ones.
\cite{Avkiran} found that collaboration leads to articles of higher impact in finance,
up to 4 collaborators.
\cite{Gazni} found a significant positive correlation between the number of authors and the number of citations in Harvard publications.
\cite{Hsu2011} considered 90k articles in natural sciences,
finding that the average number of citations
scales as $N_{\rm aut}^{1/3}$
(data extend up to about 10 co-authors), up to wide fluctuations.
\cite{Lee} found that 
the number of peer-reviewed journal papers is 
strongly and significantly associated with the number of collaborators,
unlike the number of fractionally-counted papers.
\cite{Katz} studied how the average number of citations per paper varies with different types of collaborations.
See also the works of \cite{Raan}, \cite{Sooryamoorthy} and \cite{Birnholtz} for additional studies.


General theoretical arguments concerning scale-free systems suggest that the
scientific productivity of collaborations and the corresponding frequency distribution of citations should show some, at least approximate, power law dependence on $N_{\rm aut}$. Empirical evidence appears to support these arguments.
Finding the most appropriate exponents for these scaling laws would offer the possibility of weighing the production of collaborations in the bibliometric estimate of the (quantitative) value of their results in such a way as to discourage adaptive and opportunistic behaviours while encouraging more appropriate practices in the indication of co-authorship.

In  section \ref{TH} we develop and present some theoretical arguments in favour of weighing the individual contributions to a single paper as $1/N_{\rm aut}^\alpha$, where $\alpha \approx  1/3-1/2$, with the former (latter) value more appropriate for larger (smaller) collaborations. When counting overall citations we suggest the exponent $\alpha = 1$, corresponding to fractional counting. 
By combining the two above arguments, in section \ref{hscale} we define an $h$ index that does not scale with collaboration size.

In section \ref{FC} we analyze empirical data concerning a very large number of collaborations active in fundamental physics, where the range of available values of $N_{\rm aut}$ allows for sufficiently convincing estimates of the exponents describing the dependence on $N_{\rm aut}$ of the total number of papers and of the mean and total number of citations.

Finally, we summarize and draw our conclusions in section \ref{conclusion}.

\section{A theoretical approach}\label{TH}

\subsection{Scaling}
The behaviour of collaborations with $N_{\rm aut}$ authors can be viewed as a scale-free phenomenon for a wide range of values of $N_{\rm aut}$. 
Any upper limit on $N_{\rm aut}$ would be sufficiently large to exclude any sensible effect on the equilibrium distributions in the range of values we are interested to explore (3 to 4 orders of magnitude).

We therefore expect that the various indices $N_I$ that characterise bibliometric outputs of collaborations
are distributed at equilibrium following a power-law behaviour, which we parametrise as follows
\beq \langle N_I\rangle = C_I  N^{p_I}_{\rm aut}\eeq
where $C_I$ and the powers $p_I$ are constants. 
For example this applies to the number  
of papers  produced (in a definite amount of time) by a scientific  collaboration $N_{\rm pap}=C_{\rm pap} N_{\rm aut}^{p_{\rm pap}}$
and to the average number of citations per paper $N_{\rm cit}=C_{\rm cit} N_{\rm aut}^{p_{\rm cit}}$.
The total number of citations $N_{\rm totcit}$ received by the papers of a collaboration then scales as
\beq N_{\rm totcit}=N_{\rm cit} N_{\rm pap} = 
C_{\rm cit} C_{\rm pap} N_{\rm aut}^{p_{\rm totcit}},
\qquad
p_{\rm totcit} = p_{\rm cit} + p_{\rm pap}.\eeq


\subsection{Scaling of the total number of citations}
Assuming a collective rational behaviour, and
that on average the number of citations received by scientific papers
may be considered as a reasonable proxy for their quality, 
we might expect that individual and collective choices would lead at equilibrium to 
\beq 
p_{\rm totcit}\approx 1, \eeq
namely that the total number of citations received by collaborations scales, on average,
with the number of members.
This expresses the fact that the (average) value of work made by $N_{\rm aut}$ scientists should  approximately be equal to $N_{\rm aut}$ times the work made by a single scientist. 
A lower power $p_{\rm totcit}$ would arise in the presence of gift authorships,
namely of authors who sign papers without substantially contributing.

\smallskip

This means that the total number of citations is not a fair indicator for authors, as it grows with the number of co-authors.
Similarly, the $h$-index {introduced by \cite{Hirsch:2005zc}}, being on average proportional to square root of the number of citations,
grows on average as the square root of the number of co-authors.

\medskip

A bibliometric index which does not overestimate nor underestimate
individual contributions to a collaboration 
is then the 
number of fractionally-counted citations $N_{\rm fcit}$ received by each author.
This means that a fraction $f_A$ of each paper is attributed to each co-author $A$ such that the fractions $f_A$
sum up to unity.\footnote{{We do not address the relative assignment of credit.
In some fields the contribution of different authors is reflected by their
order, with special recognition given to first and last authors.
Various proposals have been put forward to encode the relative credit in the fractions $f_A$ (see, e.g.~\cite{Kosmulski2012} and \cite{Waltman2015}).
In some other fields authors are sorted alphabetically, giving no information about who contributed more.
This happens in most papers in our data-base, so that we will assume a common $f_A$ equal to the inverse
of the number of authors.}}
On average $N_{\rm fcit}$ scales with power index $p_{\rm fcit}= p_{\rm totcit} - 1\approx 0$,
showing that $N_{\rm fcit}$ is a scale-invariant quantity that (unlike the number of citations) cannot be arbitrarily inflated grouping authors.

%

\subsection{Scaling of total number of papers}
In order to implement these concepts into actual bibliometric indices for the total number of papers
or for the average number of citations,
we must offer arguments in favour of explicit values for the exponents $p_{\rm pap}$ and $p_{\rm cit}$.\footnote{Some authors think that counting publications has no bibliometric interest, with citations being the only relevant quantity to be measured.
From such a perspective, it remains nevertheless interesting to know
how collaborations tend to split their bibliometric output within their publications.
For example one might observe a gap in citation output between (groups of) authors 
and want to understand if it mostly arises from publication intensity.
As another example, one might have a partial data-base (e.g.\ limited to some area)
that only allows to reliably compute publication intensity.
Furthermore, the publication indicator is available immediately,
while the citation indicator becomes more significant after some years as citations accumulate.
As a matter of fact, while publication intensity is a less significant indicator,
it remains used or at least mentioned because of its simplicity.
At the opposite extremum other authors think that citations can be significantly
distorted by social biases, and view publications numbers as a more objective
bibliometric indicator.
In order to obtain a result that does not scale with collaboration size, one should then fractionalize
papers according to the appropriate power for papers.}
 
We present here a simple ``theoretical'' argument. 
As the goal of collaborations is achieving more than what single authors can achieve, we expect 
$p_{\rm cit}>0$.
Assuming rational behaviour in the formation of collaborations one may expect that (at least for not-too-big groups), individual competence of partners be as far as possible complementary, and therefore ``orthogonal'' in some abstract 
$N$-dimensional ``space of competencies''. 
We may therefore regard the qualitative output of a collaboration as the vector sum of $N$ orthogonal vectors.

\begin{figure}[t]
$$\includegraphics[width=0.36\textwidth]{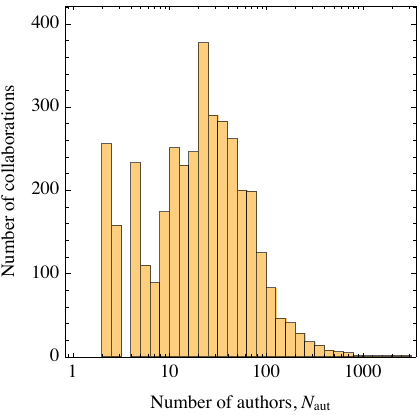}\qquad
\includegraphics[width=0.52\textwidth]{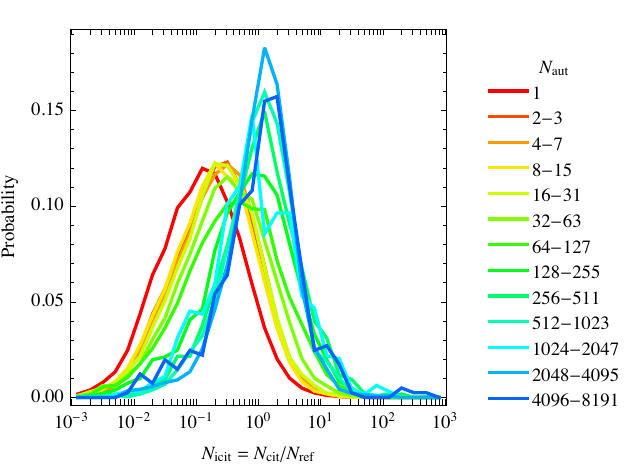}
$$
\caption{\label{fig:NcolNaut} 
{\bf Left}:
Total number of collaborations listed in the \inspire database with the number of authors shown on the horizontal axis.
{\bf Right}: distribution of the number of individual citations 
(citations divided by the number of references of the citing papers)
received by papers with the indicated number of authors.
}
\end{figure}

The limit where all authors have ``orthogonal'' competencies corresponds to
$N= N_{\rm aut}$: then the length of such vector scales as $N_{\rm aut}^{1/2}$.\footnote{Assuming
that authors have different skills, the average length squared of such vector scales as $N_{\rm aut}$.}
This corresponds to
\beq \label{eq:1/2}
p_{\rm pap}=1/2,\qquad p_{\rm cit}=1/2.\eeq
Sometimes, more than one collaborator is needed to fulfil a needed competency:
realistic collaborations organise in $N_{\rm sub} \le N_{\rm aut}$ sub-collaborations 
that work on ``orthogonal'' topics.
We assume the number of sub-collaborations satisfies the
scaling law 
\beq
N_{\rm sub} = N_{\rm aut}^s\qquad  \hbox{with exponent $s\le 1$.}\eeq
Then, the average number of citations of each paper scales as
the square root of the number of ``orthogonal'' competencies $N = N_{\rm sub}$:
\beq  N_{\rm cit}\propto \sqrt{N} \propto N_{\rm aut}^{s/2}. \eeq
Assuming again an optimal distribution of resources, 
$N_{\rm totcit}$ is expected to scale as $N_{\rm aut}$,
and thereby the number of papers is expected to scale as
\beq N_{\rm pap} \propto N_{\rm aut}^{1-s/2}.
\eeq
It is reasonable to assume that the number of papers 
scales as $N$, the number of topics about which the collaboration has competencies.
This leads to $s=1-s/2$, solved by $s=2/3$, and thereby to
\beq p_{\rm pap}=2/3,\qquad p_{\rm cit}=1/3.\eeq
A weaker growth of the number of papers with $N$ leads to smaller $p_{\rm pap}$.

%
%

\subsection{Scaling of the $h$ index}\label{hscale}
The $h$ index (defined by~\cite{Hirsch:2005zc} as the number of papers that received more than $h$ citations),
provides extra information on the distribution of the number of citations, 
favouring authors that produced many highly-cited papers with respect to authors
that produced many poorly cited papers plus a small number of top-cited papers.

Theoretical arguments (see e.g.~\cite{Yong2014}) and evidence from data analysis  (see e.g.~\cite{MannellaRossi2013} and \cite{Strumia:2018zgi}) indicate that the $h$ index is strongly correlated to the square root of the total number of citations received by an author: 
\beq h \approx \alpha \,N_{\rm totcit}^{0.5}.\eeq
where the theoretical prediction from \cite{Yong2014} is $\alpha \approx 0.54$ and the phenomenological result  obtained from the data of about 1400 Italian physicists is $\alpha \approx 0.53$~\citep{MannellaRossi2013}.

Like the number of citations, the $h$ index is affected by the collaboration size,
being higher for authors with more collaborators.
Assuming for the $h$ index the above mentioned scaling as a function of $N_{\rm totcit}$, and 
recalling our prediction $p_{\rm totcit}\approx 1,$
we thereby expect that  the $h$-index should scale approximately as the square root of the (average) number of authors:
\beq h \propto \,N_{\rm aut}^{0.5}\eeq
independently of the specific values taken by $p_{\rm pap}$ and $p_{\rm cit}$, as long as they satisfy $p_{\rm cit} + p_{\rm pap} \approx 1$.

\section{Data about collaborations in fundamental physics}\label{FC}
In this section  we  present bibliometric data in  fundamental physics,
that offer support for 
\beq p_{\rm pap}\approx 0.5-0.6,\qquad 
p_{\rm cit} \approx 0.4-0.5,\qquad
p_{\rm totcit}\approx 1,\qquad
p_{\rm fcit}\approx 0.\eeq
We use the  {\sc InSpire}\footnote{High-Energy Physics Literature \inspire Database (\href{https://inspirehep.net/}{https://inspirehep.net}).} database
that gives 
a picture of fundamental physics world-wide from $\sim1970$ to end 2019:
1.34 millions of scientific
papers, 32 millions of references, 75 thousands of identified authors.
Fundamental physics contains large collaborations, up to 3000 authors
that produced 6000 publications.  Adopting full counting these are counted as
$3000\times 6000$ publications, dominating the whole database
and producing bibliometric indices uncorrelated to human
evaluations of scientific merit.
Fundamental physics thereby is a good sample to study how the bibliometric outputs of collaborations scale
with the number of collaborators.
We will show data for two different kinds of collaborations:
\begin{enumerate}
\item {\bf Official collaborations}.
We consider the 5965 (mostly experimental) collaborations listed in the \inspire database.
Each collaboration produced a certain number $N_{\rm pap}$
of papers, roughly written with the same group of $N_{\rm aut}$ authors.
The left panel of Figure\fig{NcolNaut} gives some demographic information, that is the distribution of the number of authors in collaborations
and the distribution of fractionally counted citations for different collaboration sizes. In the following, when showing results for
official collaborations, we indicate collaborations as dots in a scatter plot, with the main collaborations indicated by their names.
Furthermore, we show the mean (median) as a red (magenta) curve and a blue dotted line highlighting the scaling with the number of authors.\footnote{{A few collaborations
varied significantly their number of authors.
We define the number of authors of a collaboration
by averaging the number of authors of all its papers, with weights proportional to their number of citations.
This procedure assigns minor weight to proceedings written by one or few authors and to papers
written by earlier incomplete phases of the collaboration.}}

\item {\bf Occasional collaborations}.
Many more multi-authored papers have been
written by collaborations that form for one or few papers.
To study them we proceed as follows.
For each author in the \inspire database we compute the average number of authors of
his/her papers, $\langle N_{\rm aut}\rangle\ge1$, as well as his/her bibliometric indices
(number of papers, of citations, etc).
In view of the large number of authors, in the following, when presenting results on occasional collaborations, 
we avoid showing scatter plots and only show averages. Moreover, results are shown separately within the main topics of
fundamental physics: experiment, theory, astro/cosmo.
The first category includes all papers in the
hep-ex (high-energy experiments) and nucl-ex (nuclear experiments)
category of arXiv.
The latter category includes papers in 
astro-ph, which contains astrophysics and cosmology.
Theoretical papers are those appeared in 
hep-ph (high-energy phenomenology), hep-th (high-energy theory),
hep-lat (lattice), nucl-th (nuclear theory),
gr-qc (general relativity and quantum cosmology).
\end{enumerate}

\begin{figure}[t]
$$\includegraphics[width=0.4\textwidth]{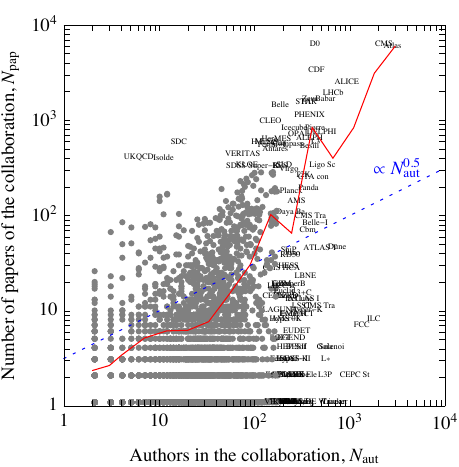}\qquad
\includegraphics[width=0.4\textwidth]{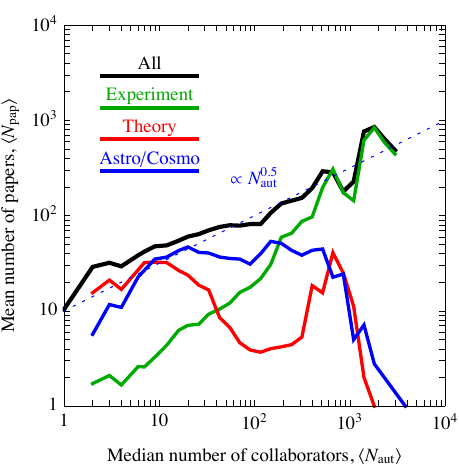}
$$
\caption{\label{fig:NautNpap} 
Number of papers versus number of collaborators. 
}
\end{figure}

\subsection{Scaling of the number of papers}
Figure\fig{NautNpap} shows that
the number of  papers produced by
official (left panel) or occasional (right panel) collaborations
scales with the number of authors as
\beq \label{eq:Npap} N_{\rm pap} \propto N_{\rm aut}^{0.5-0.6}.\eeq
In the right panel, theoretical papers 
with many authors fall below the scaling.
These are rare outliers:  almost all papers in theoretical categories have few authors.
Theoretical papers with many authors mostly are collections of separate contributions
grouped together, rather than big collaborations.

\subsection{Scaling of the mean number of citations}
\begin{figure}[t]
$$\includegraphics[width=0.4\textwidth]{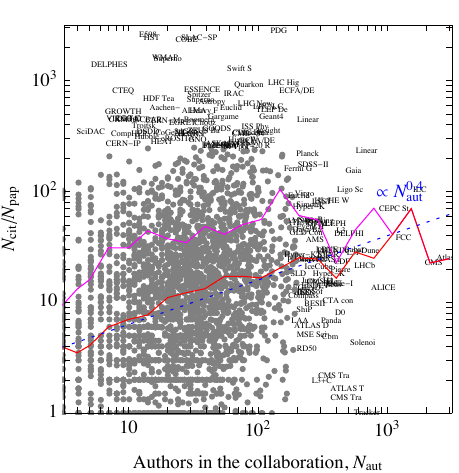}\qquad
\includegraphics[width=0.4\textwidth]{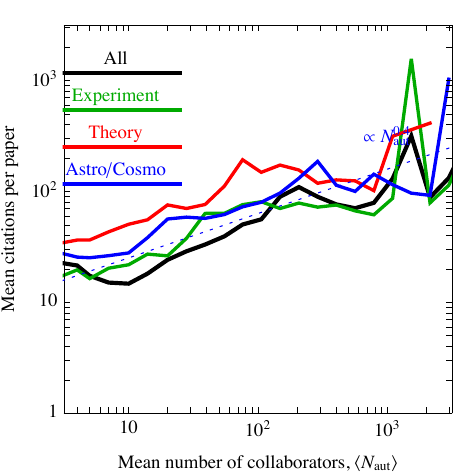}
$$
\caption{\label{fig:MeanNcitNaut} 
Mean number of citations per paper versus number of collaborators. }
\end{figure}

Figure\fig{MeanNcitNaut} shows that the mean number of citations received by 
papers written by an official collaboration
(left panel) or by an author (right panel)
roughly scales with the average number of co-authors as
\beq \label{eq:Ncit}
N_{\rm cit}=
\frac{N_{\rm totcit}}{N_{\rm pap}} \propto N_{\rm aut}^{0.4-0.5}.\eeq
This result is in reasonable agreement with~\cite{Hsu2011}, who found, in a much smaller sample,
a power $\approx 1/3$ up to $N_{\rm aut}\sim 10$.

\begin{figure}[t]
$$\includegraphics[width=0.4\textwidth]{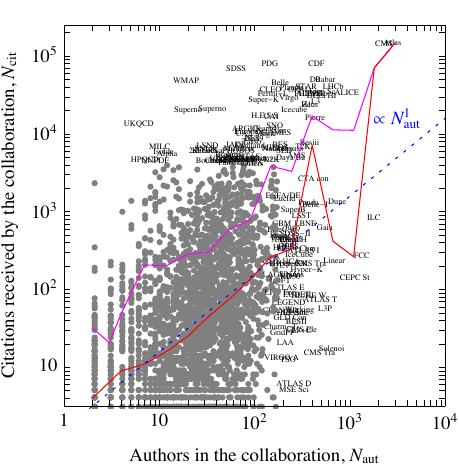}\qquad
\includegraphics[width=0.4\textwidth]{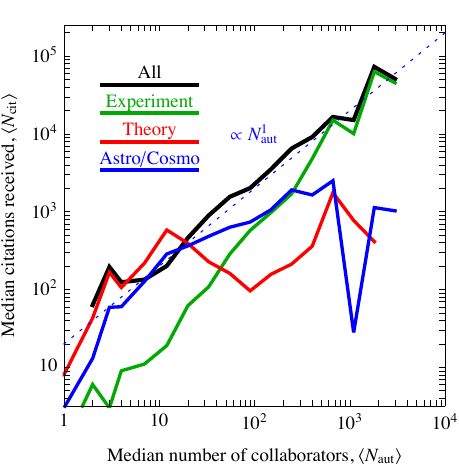}
$$
\caption{\label{fig:NautNcit} 
Number of citations versus number of collaborators. }
\end{figure}
\begin{figure}[t!]
$$\includegraphics[width=0.4\textwidth]{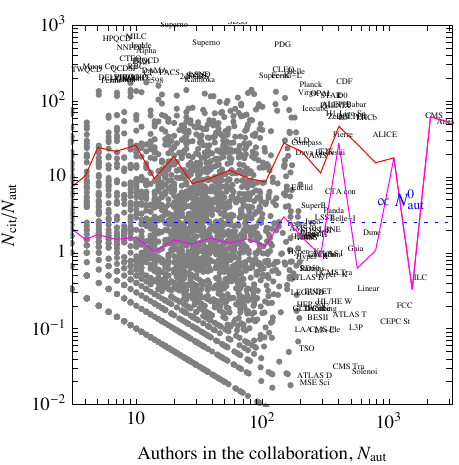}\qquad
\includegraphics[width=0.4\textwidth]{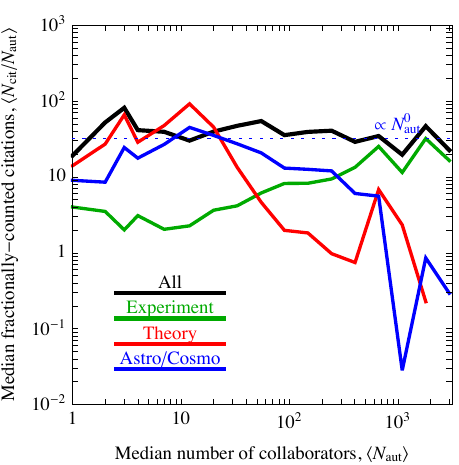}$$
\caption{\label{fig:NautNfcit} 
Total number of fractionally-counted citations versus number of collaborators. }
\end{figure}

%
%
%
%
%

\subsection{Scaling of the total number of citations}
Figure\fig{NautNcit} shows that the total number of citations received by an official collaboration
(left panel) or  author (right panel)
grows roughly linearly with the average number of co-authors:
\beq N_{\rm totcit} \propto N_{\rm aut}^{1}.\eeq
This is expected by combining the two previous scalings:
the total number of citations of a collaboration
can be decomposed as the product of
the number of papers written by the collaboration, 
times the average number of citations received per paper:
these factors roughly scale as
$N^{0.5-0.6}_{\rm aut} \times N_{\rm aut}^{0.4-0.5}$.

\medskip

Figure\fig{NautNfcit} shows that the  total number of fractionally-counted citations
$N_{\rm fcit} = \sum_p N_{p\rm cit}/N_{p\rm aut}$
received by papers $p$ written by an official collaboration
(left panel) or author (middle panel) is roughy independent of the average number of  co-authors.
A similar result holds for a related quantity, ``individual citations'', defined as fractionally counted
citations divided by the number of references of the citing papers:
 \beq  N_{\rm fcit},N_{\rm icit} \propto N_{\rm aut}^{0}.\eeq
This means that fractionally-counted citations or individual citations neither
reward nor penalise working in big collaborations,
while citations reward authors who prefer working in
big collaborations.

\subsection{Scaling of the total number of field-normalized citations}
When analysing different fields it is usually meaningful to correct for their
different publication intensities.  As different fields also show different collaboration patterns,
this opens the issue about how to properly account for the two aspects in a combined way.
A simple general answer is obtained by counting citations divided by the number of references
of the citing paper, namely $N_{\rm icit} \equiv N_{\rm cit}/N_{\rm aut}^{p_{\rm totcit}}N_{\rm ref}$
(for a precise definition see~\cite{Strumia:2018zgi} where this indicator is dubbed ``individual citations'').
This quantity is similar to citations (so that $p_{\rm totcit}\approx 1$) and it automatically
provides a field-independent indicator, without having to identify fields.
Indeed, within any hypothetical closed field (with no citations to or from other fields)
this indicator satisfies the sum rule $\sum N_{\rm icit}= N_{\rm pap}$,
up to recent papers that will receive citations in the future.
In words, papers in sectors with higher publication intensity tend to receive more citations and thereby 
also tend to have more references.
One can thereby use references to factor out publication intensity
(see e.g.~\cite{Zitt2008} and \cite{Waltman2015} for a review).

\begin{figure}[t]
$$\includegraphics[width=0.4\textwidth]{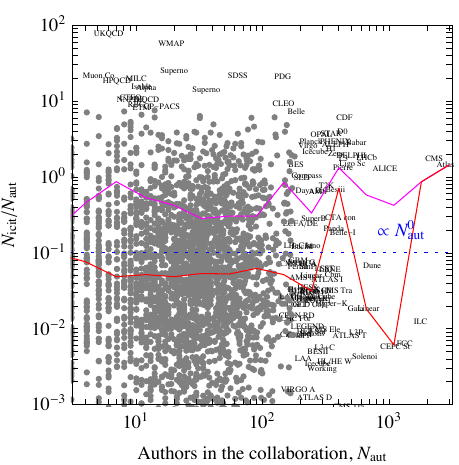}\qquad
\includegraphics[width=0.4\textwidth]{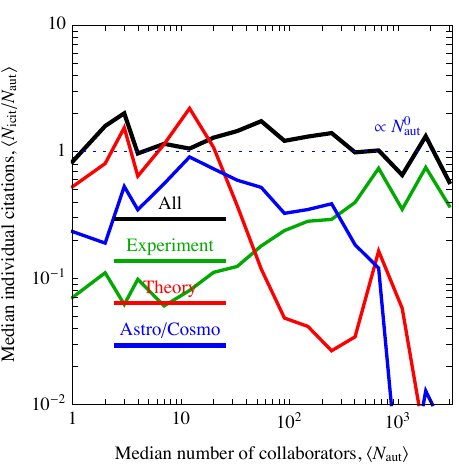}$$
\caption{\label{fig:NautNicit} 
Total number of individual citations (fractionally-counted citations divided by
references of citing papers) versus number of collaborators. }
\end{figure}

Figure\fig{NautNicit} shows this field-normalised indicator applied to fundamental physics,
showing that it exhibits the desired negligible scaling with collaboration size.

\begin{figure}[t!]
$$\includegraphics[width=0.4\textwidth]{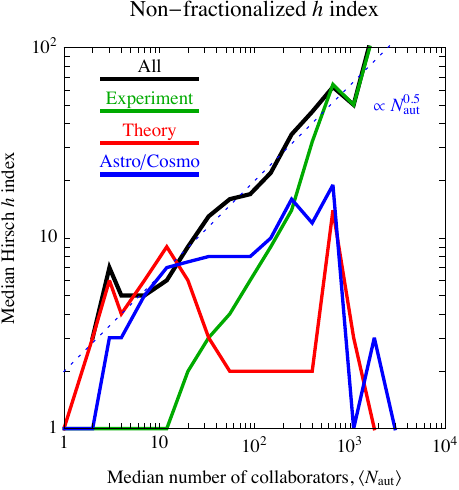}\qquad
\includegraphics[width=0.407\textwidth]{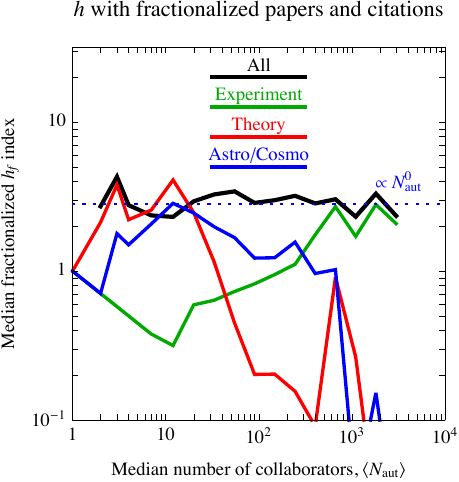}$$
$$\includegraphics[width=0.409\textwidth]{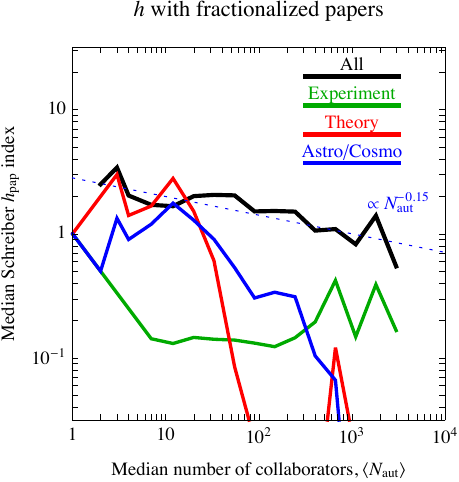}\qquad
\includegraphics[width=0.4\textwidth]{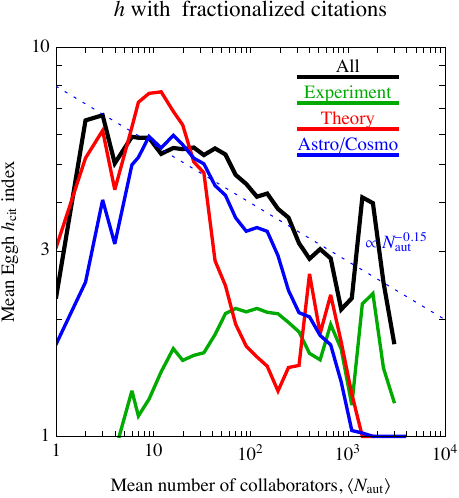}$$
\caption{\label{fig:hhm} {\bf Top-left:} the $h$-index scales as the square
root of the average number of collaborators.
{\bf Top-right}: the $h$-index fractionalized as suggested in eq.\eq{hf}
does not scale with the average number of collaborators.
{\bf Bottom}: the $h$-index fractionalized with respect to papers (left)
or to citations (right)
shows a reduced residual scaling with the average number of collaborators. 
}
\end{figure}
\subsection{Scaling of the (fractionalized) Hirsch $h$ index}
%
Looking at data, the top-left panel of fig.\fig{hhm} shows that 
the $h$ index scales with  the number of authors as
\beq h \propto N_{\rm aut}^{0.5-0.6}\eeq
consistently with our expectation.
In order to avoid this scaling, various authors defined modified $h$ indices
with fractionalized counting.
As the $h$ index combines  information on the number of citations and of papers,
one can fractionalize  with respect to citations and/or with respect to papers.
The first option was considered by \cite{Egghe08a}, who defined an index (here called $h_{\rm cit}$)
equal to the number of papers that received more than $h_{\rm cit}$ fractionalized citations.
The second option was considered by \cite{Schreiber2008}, who defined an index (here called $h_{\rm pap}$)
equal to the number of fractionally-counted papers that received more than $h_{\rm pap}$ citations.
Looking at data, the lower row of fig.\fig{hhm} shows that the these two possibilities
are only partially
successful, as they both still exhibit a milder scaling with the average number of co-authors
\beq h_{\rm cit} ,h_{\rm pap}\propto N_{\rm aut}^{-0.15}.\eeq
In order to understand the reason of this residual scaling with the average number of co-authors,
we recall that larger collaborations tend to produce more papers as well as papers that get more cited.
The $h$ index fractionalized  with respect to papers
would exhibit the desired scale-invariant behaviour if
the scalings were $N_{\rm pap}\propto N_{\rm aut}^{p_{\rm pap}}$ and
$N_{\rm cit}\propto N_{\rm aut}^{p_{\rm cit}}$
with $p_{\rm pap}=1$ and $p_{\rm cit}=0$.
Similarly, the $h$ index fractionalized  with respect to citations
would exhibit the desired scale-invariant behaviour if
$p_{\rm pap}=0$ and $p_{\rm cit}=1$.

Instead, data  show  $p_{\rm pap}\approx p_{\rm cit}\approx 1/2$
(see eq.\eq{Npap}  and eq.\eq{Ncit}), in agreement with our model values of eq.\eq{1/2}.
Thereby a fractionalized $h$-index $h_f$ that exhibits the desired scale-invariant behaviour
is obtained by partially fractionalizing with respect to both papers and citations.
To compute our $h_f$ one needs to
sort papers according to citations partially fractionalized as $c=N_{\rm cit}/\sqrt{N_{\rm aut}}$,
and summing authorships partially fractionalized as $1/\sqrt{N_{\rm aut}}$ 
until $h_f$ is larger than the fractionalized $c$.
In formul\ae
\beq \label{eq:hf} h_f = \sum_{h_f < N_{\rm cit}/\sqrt{N_{\rm aut}}}  \frac{1}{\sqrt{N_{\rm aut}}}\eeq
where the  sum runs over all papers of the considered author.
The top-right panel of fig.\fig{hhm} shows that $h_f$ exhibits the expected scale-invariant behaviour,
and can thereby be used to quantify the scientific output of authors in a way that does not
depend on their average number of collaborators.

Another simpler index  that achieves the same independence 
is the number of fractionally counted citations $N_{\rm fcit}$ plotted in fig.\fig{NautNfcit}.
The $h_f$ index, analogously to the $h$ index, provides extra information on the distribution
of the number of citations, and has the same intuitive meaning as the Schreiber $h_m$ index, that here we called $h_{\rm pap}$.


%

\section{Summary and conclusions}\label{conclusion}
In this paper we studied the bibliometric output of collaborations.
Our study was motivated by the present situation in fundamental physics, where
collaborations can be so large that accounting for their size
has a huge bibliometric impact. Nevertheless, our results also apply to fields with smaller collaborations, 
up to the quantitative difference that the factors we considered have less numerical relevance.

Clearly, some partial fractional counting of bibliometric quantities is the solution to the problem. 
However, unless the correct amount of fractional counting appropriate to any given  bibliometric indicator
is used, the index will favour or disfavour scientists that choose to carry out their research in small or large groups.
This is a particularly relevant matter, given that there are nowadays several policymakers that implement bibliometric evaluation in their funding and hiring criteria.

The field of fundamental research represents the natural playground to study bibliometrics for collaborations, 
both in terms of the large available dataset provided by \inspire and in terms of the large variability in collaboration 
size, that is by far the largest in the public research domain, reaching thousands of authors.
To our knowledge there are no other studies on this topic carried out with such a dataset, 
so that our results cannot be confronted with previous studies.

Having to deal with $N_{\rm aut}\gg 1$ authors, we started
in section \ref{TH} with general theoretical considerations on the behaviour of scientific research
that lead us to a series of assumptions and to the subsequent
formulation of hypotheses on the scaling of bibliometrics indicators with the number of authors. 
The crucial hypotheses are that bibliometric quantities for collaboration work exhibit a power-law behaviour 
with the number of authors in the collaboration, and that, 
in a situation of ``equilibrium'' the scaling of the total number of citations received by a collaboration, summing up all its papers, scales linearly with the number of authors. 
Moreover, in the hypothetical situation in which all authors contribute with fully ``orthogonal'' competences, the scaling of the number of papers
and of the number of citations per paper are both equal to $1/2$.
We therefore formulated a theoretical model for bibliometrics of collaborations based on the aforementioned considerations and assumptions. As any model,
it needs to be confronted with data to extract unknowns, that, in our case are the scaling exponents of the power-law dependence of the number of papers $p_{\rm pap}$,
the number of citations per paper $p_{\rm cit}$, the total number of citations $p_{\rm totcit}$, and the number of fractionally counted citations $p_{\rm fcit}$.

In section \ref{FC} we computed all the quantities relevant for our model and extracted the unknowns from the data in the \inspire dataset. On the one hand we observed in data the expected approximate power-law scaling. 
On the other hand, we were able to estimate the desired exponents, observing $p_{\rm pap}\approx 0.5-0.6$,
$p_{\rm cit} \approx 0.4-0.5$, $p_{\rm totcit}\approx 1$, and $p_{\rm fcit}\approx 0$.

Incidentally, given the relatively common use of the $h$ index, despite its well known correlation with the number of citations, 
we also predicted and evaluated the scaling with the number of authors of the $h$ index, $ h \propto N_{\rm aut}^{0.5-0.6}$.
Furthermore, we defined a modified $h$ index (see eq.\eq{hf}) that roughly does not scale with the number of authors. 
We confronted our result with different modified $h$ indices proposed in the literature, which did not solve the issue of dependence on the number of authors.

Our results apply to mean (or median) quantities. Before concluding, a comment on the full distributions, or at least on their variances is in order. In the right panel of Figure\fig{NcolNaut} we show the distributions
of individual citations received by all papers in our database, splitting them according to their number of authors. We see that, as already suggested in the literature, 
papers with more authors are more cited. We also see that distributions have large variabilities: the distributions are approximatively log-normal with log-scale means that scale as 
$\langle N_{\rm cit}\rangle \propto {N_{\rm aut}^{0.5}} $ and with log-scale widths that remain approximatively constant.
This behaviour is obtained from our initial theoretical considerations adding one extra assumption: that collaborations tend to equalise the total amount of skill within each competence,
such that the distribution in $N_{\rm cit}$ of a collaboration is simply obtained rescaling the distribution of single-author papers.
The bibliometric output of collaborations formed as random groups of authors would instead show larger variabilities.

\small



\normalsize

\bibliographystyle{apacite}
\bibliography{artColl}

\end{document}